\documentclass{PoS}
\usepackage{amsmath}

\title{
\begin{flushright}
 { \small CERN-PH-TH/2012-079}
\end{flushright} 
\vskip 1.5cm
Dimensional Reduction of ${\cal N}=1$, $E_8$ SYM over  {
$SU(3)/U(1)\times U(1)\times \mathbb{Z}_3$} and its four-dimensional effective
action}

\ShortTitle{Dimensional Reduction}

\author{\speaker{Nikos Irges} \\
National Technical University of Athens 
Zografou Campus, GR-15780 Athens Greece  \\
 E-mail: \email{irges@mail.ntua.gr}  
}

\author{Georgios Orfanidis\thanks{On leave from Physics Department, 
National Technical University, Zografou Campus:
Heroon Polytechniou 9, 15780 Zografou, Athens, Greece}\\
        Theory Group, Physics Department CERN, Geneva, Switzerland\\
        E-mail: \email{orfanidi@mail.ntua.gr}}

\author{George Zoupanos \footnotemark[\value{footnote}] \\
        Theory Group, Physics Department CERN, Geneva, Switzerland\\
        E-mail: \email{George.Zoupanos@cern.ch}}

\abstract{
We present an extension  of the Standard Model inspired by the 
$E_8\times E_8$ Heterotic String. In order that a reasonable effective Lagrangian is presented
we neglect everything else other than the ten-dimensional ${\cal N}=1$ supersymmetric Yang-Mills
sector associated with one of the gauge factors and certain couplings necessary for anomaly 
cancellation. 
We consider a compactified space-time 
$M_4 \times B_0/\mathbb{Z}_3$, where $B_0$ is the nearly-K\"ahler manifold
$SU(3)/U(1) \times U(1)$ and $\mathbb{Z}_3$ is a freely acting discrete group on
$B_0$.
Then we reduce dimensionally the $E_8$ on this manifold and we employ the Wilson flux mechanism 
leading in four dimensions to an $SU(3)^3$ gauge theory with the spectrum of a ${\cal N}=1$ supersymmetric theory.
We compute the effective four-dimensional Lagrangian and demonstrate that an
extension of the Standard Model is obtained with interesting features including a
conserved baryon number and fixed tree level Yukawa couplings and scalar potential.
The spectrum contains new states such as right handed neutrinos and heavy
vector-like quarks.
}

\FullConference{Proceedings of the Corfu Summer Institute 2011 School and Workshops on Elementary Particle Physics and Gravity\\
		September 4-18, 2011\\
		Corfu, Greece}

\begin{document}
\section{Introduction}
Superstring Theory is often regarded as the best candidate for a quantum theory
of gravitation, or more generally as a unified theory of all fundamental
interactions. On the other hand the main goal expected from a unified
description of interactions by the Particle Physics community is to understand
the present day large number of free parameters of the Standard Model (SM) in
terms of a few fundamental ones. Indeed the celebrated SM had so far outstanding
successes in all its confrontations with experimental results. However its
apparent success is spoiled by the presence of a plethora of free parameters
mostly related to the ad-hoc introduction of the Higgs and Yukawa sectors in the
theory.

 It is worth recalling that various dimensional reduction schemes, with the
Coset Space Dimensional Reduction (CSDR) \cite{Manton,Review,Kuby} and the
Scherk-Schwarz reduction \cite{Scherk} being pioneers, suggest that a
unification
of the gauge and Higgs sectors can be achieved in higher than four dimensions.
The four-dimensional gauge and Higgs fields are simply the surviving components
of the gauge fields of a pure gauge theory defined in higher dimensions, while
the addition of fermions in the higher-dimensional gauge theory leads naturally
after CSDR to Yukawa couplings in four dimensions. The last step in this unified
description in high dimensions is to relate the gauge and fermion fields, which
can be achieved by demanding that the higher-dimensional gauge theory is ${\cal
N}=1$ supersymmetric, i.e. the gauge and fermion fields are members of the same
vector supermultiplet. Furthermore a very welcome additional input coming from
Superstring Theory (for instance the heterotic string \cite{Gross1}) is the
suggestion of the space-time dimensions and the gauge group of the
higher-dimensional supersymmetric theory \cite{Theisen}.

Superstring Theory is consistent only in ten dimensions and therefore the
following crucial issues have to be addressed, (a) distinguish the extra
dimensions from the four observable ones which are experimentally approachable,
i.e. determine a suitable compactification which is a solution of the theory (b)
reduce the higher-dimensional theory to four dimensions and determine the
corresponding four-dimensional theory, which may subsequently be compared to
observations. Among superstring theories the heterotic string \cite{Gross1} has
always been considered as the most promising version in the prospect to find
contact with low-energy physics studied in accelerators, mainly due to the
presence of the ten-dimensional ${\cal N}=1$ gauge sector. Upon compactification
of the ten-dimensional space-time and subsequent dimensional reduction the
initial $E_8 \times E_8$ gauge theory can break to phenomenologically
interesting Grand Unified Theories (GUTs), where the SM could in principle be
accommodated \cite{Gross2}. Moreover, the presence of chiral fermions in the
higher-dimensional theory serves as an advantage in view of the possibility to
obtain chiral fermions also in the four-dimensional theory. Finally, the
original ${\cal N}=1$ supersymmetry can survive and not get enhanced in four
dimensions, provided that appropriate compactification manifolds are used. In
order to find contact with the minimal supersymmetric standard model (MSSM), the
non-trivial part of this scenario was to invent mechanisms of supersymmetry
breaking within the string framework.

The task of providing a suitable compactification and reduction scheme which
would lead to a realistic four-dimensional theory has been pursued in many
diverse ways for more than twenty years. The realization that Calabi-Yau (CY)
threefolds serve as suitable compact internal spaces in order to maintain an
${\cal N}=1$ supersymmetry after dimensional reduction from ten dimensions to
four \cite{Candelas} led from the beginning to pioneering studies in the
dimensional reduction of superstring models \cite{Witten2,Derendinger}. However,
in CY compactifications the resulting low-energy field theory in four dimensions
contains a number of massless chiral fields, known as moduli, which correspond
to flat directions of the effective potential and therefore their values are
left undetermined. The attempts to resolve the moduli stabilization problem led
to the study of compactifications with fluxes (for reviews see e.g.
\cite{Grana}). In the context of flux compactifications the recent developments
suggested the use of a wider class of internal spaces, called manifolds with
$SU(3)$-structure, which contains CYs. Admittance of an $SU(3)$-structure is a
milder condition as compared to $SU(3)$-holonomy, which is the case for CY
manifolds, in the sense that a nowhere-vanishing, globally-defined spinor can be
defined such that it is covariantly constant with respect to a connection with
torsion and not with respect to the Levi-Civita connection as in the CY case. An
interesting class of manifolds admitting an $SU(3)$-structure is that of
nearly-K\"ahler manifolds. The homogeneous nearly-K\"ahler manifolds in six
dimensions have been classified in \cite{Butruille} and they are the three
non-symmetric six-dimensional coset spaces (see table \ref{table4} Appendix
\ref{tables}) and the group manifold $SU(2) \times
SU(2)$. In the studies of heterotic compactifications the use of non-symmetric
coset spaces was introduced in \cite{Govindarajan1,DLust,Govindarajan2} and
recently developed further in \cite{Lopes,MPZ}. Particularly, in \cite{MPZ} it
was shown that supersymmetric compactifications of the heterotic string theory
of the form $AdS_4 \times S/R$ exist when background fluxes and general
condensates are present. Moreover, the effective theories resulting from
dimensional reduction of the heterotic string over nearly-K\"ahler manifolds
were
studied in \cite{CMZ}.

Last but not least it is worth noting that the dimensional reduction of
ten-dimensional ${\cal N}=1$ supersymmetric gauge theories over non-symmetric
coset spaces led in four dimensions to softly broken ${\cal N}=1$ theories
\cite{Pman,Pman2}.

Here we would like to present the significant progress that has been made
recently concerning the dimensional reduction of the N=1 supersymmetric $E_8$
gauge
theory resulting in the field theory limit of the heterotic string over the
nearly-K\"ahler manifold $SU(3)/U(1) \times U(1)$. Specifically an extension of
the Standard Model (SM) inspired by the $E_8 \times E_8$ heterotic string was
derived \cite{Irges:2011de}\footnote{For earlier attempts to obtain realistic
models by CSDR see ref
\cite{Review,Chapline,Kapetanakis,Hanlon}.}. In addition in
order to make the presentation
self-contained we
present first a short review of the CSDR.

\section{The Coset Space Dimensional Reduction.}
Given a gauge theory defined in higher dimensions the obvious way
to dimensionally reduce it is to demand that the field dependence
on the extra coordinates is such that the Lagrangian is
independent of them. A crude way to fulfill this requirement is to
discard the field dependence on the extra coordinates, while an
elegant one is to allow for a non-trivial dependence on them, but
impose the condition that a symmetry transformation by an element
of the isometry group $S$ of the space formed by the extra
dimensions $B$ corresponds to a gauge transformation. Then the
Lagrangian will be independent of the extra coordinates just
because it is gauge invariant. This is the basis of the CSDR
scheme \cite{Manton,Review,Kuby}, which assumes that $B$ is a
compact coset space, $S/R$.

In the CSDR scheme one starts with a Yang-Mills-Dirac Lagrangian,
with gauge group $G$, defined on a
 $D$-dimensional spacetime $M^{D}$, with metric $g^{MN}$, which is compactified
to $ M^{4}
\times S/R$ with $S/R$ a coset space. The metric is assumed to
have the form
\begin{equation}\label{pan:1}
g^{MN}=
\left[\begin{array}{cc}\eta^{\mu\nu}&0\\0&-g^{ab}\end{array}
\right],
\end{equation}
where $\eta^{\mu\nu}= diag(1,-1,-1,-1)$ and $g^{ab}$ is the coset
space metric. The requirement that transformations of the fields
under the action of the symmetry group of $S/R$ are compensated by
gauge transformations lead to certain constraints on the fields.
The solution of these constraints provides us with the
four-dimensional unconstrained fields as well as with the gauge
invariance that remains in the theory after dimensional reduction.
Therefore a potential unification of all low energy interactions,
gauge, Yukawa and Higgs is achieved, which was the first
motivation of this framework.

It is interesting to note that the fields obtained using the CSDR
approach are the first terms in the expansion of the
$D$-dimensional fields in harmonics of the internal space $B$
\cite{Review,Salam}. The
effective field theories resulting from compactification of higher
dimensional theories might contain also towers of massive higher
harmonics (Kaluza-Klein) excitations, whose contributions at the
quantum level alter the behaviour of the running couplings from
logarithmic to power \cite{Taylor}. As a result the traditional
picture of unification of couplings may change drastically
\cite{Dienes}. Higher dimensional theories have also been studied
at the quantum level using the continuous Wilson renormalization
group \cite{Kubo} which can be formulated in any number of
space-time dimensions with results in agreement with the treatment
involving massive Kaluza-Klein excitations. However we should stress that in
ref \cite{Chatzistavrakidis:2007by} the CSDR has been shown to be a consistent
scheme.

Before we proceed with the description of the  CSDR scheme we need
to recall some facts about coset space geometry needed for
subsequent discussions. Complete reviews can be found in
\cite{Review,Castellani}.
\subsection{Coset Space Geometry.}
Assuming a $D$-dimensional spacetime $M^{D}$ with metric $g^{MN}$
given in \eqref{pan:1} it is instructive to explore further the geometry
of all coset spaces $S/R$.

We can divide the generators of $S$, $ Q_{A}$ in two sets : the
generators of $R$, $Q_{i}$ $(i=1, \ldots,dimR)$, and the
generators of $S/R$, $ Q_{a}$( $a=dimR+1 \ldots,dimS)$, and
$dimS/R=dimS-dimR =d$. Then the commutation relations for the
generators of $S$ are the following:
\begin{eqnarray}\label{pan:2}
\left[ Q_{i},Q_{j} \right] &=& f_{ij}^{\ \ k} Q_{k},\nonumber \\
\left[ Q_{i},Q_{a} \right]&=& f_{ia}^{\ \ b}Q_{b},\nonumber\\
\left[ Q_{a},Q_{b} \right]&=& f_{ab}^{\ \ i}Q_{i}+f_{ab}^{\ \
c}Q_{c} .
\end{eqnarray}
So $S/R$ is assumed to be a reductive but in general non-symmetric
coset space. When $S/R$ is symmetric, the $f_{ab}^{\ \ c}$ in
\eqref{pan:2} vanish. Let us call the coordinates of $M^{4} \times S/R$
space $z^{M}= (x^{m},y^{\alpha})$, where $\alpha$ is a curved
index of the coset,  $a$ is a tangent space index and $y$ defines
an element of $S$ which is a coset representative, $L(y)$. The
vielbein and the $R$-connection  are defined through the
Maurer-Cartan form which takes values in the Lie algebra of $S$ :
\begin{equation}\label{pan:3}
L^{-1}(y)dL(y) = e^{A}_{\alpha}Q_{A}dy^{\alpha} .
\end{equation}
Using \eqref{pan:3} we can compute that at the origin $y = 0$, $
e^{a}_{\alpha} = \delta^{a}_{\alpha}$ and $e^{i}_{\alpha} = 0$. A
connection on $S/R$ which is described by a connection-form
$\theta^{a}_{\ b}$, has in general torsion and curvature. In the
general case where torsion may be non-zero, we calculate first the
torsionless part $\omega^{a}_{\ b}$ by setting the torsion form
$T^{a}$ equal to zero,
\begin{equation}
T^{a} = de^{a} + \omega^{a}_{\ b} \wedge e^{b} = 0,
\end{equation}
while using the Maurer-Cartan equation,
\begin{equation}
de^{a} = \frac{1}{2}f^{a}_{\ bc}e^{b}\wedge e^{c} +f^{a}_{\
bi}e^{b}\wedge e^{i},
\end{equation}
we see that the condition of having vanishing torsion is solved by
\begin{equation}
\omega^{a}_{\ b}= -f^{a}_{\ ib}e^{i}-\frac{1}{2}f^{a}_{\ bc}e^{c}-
\frac{1}{2}K^{a}_{\ bc}e^{c},
\end{equation}
where $K^{a}_{\ bc}$ is symmetric in the indices $b,c$, therefore
$K^{a}_{\ bc}e^{c} \wedge e^{b}=0$. The $K^{a}_{\ bc}$ can be
found from the antisymmetry of $\omega^{a}_{\ b}$, $\omega^{a}_{\
b}g^{cb}=-\omega^{b}_{\ c}g^{ca}$, leading to
\begin{equation}
K_{\ \ bc}^{a}=g^{ad}(g_{be}f_{dc}^{\ \ e}+g_{ce}f_{db}^{\ \ e}).
\end{equation}
In turn $\omega^{a}_{\ b}$ becomes
\begin{equation}
\omega^{a}_{\ b}= -f^{a}_{\ ib}e^{i}-D_{\ \ bc}^{a}e^{c},
\end{equation}
where $$ D_{\ \ bc}^{a}=\frac{1}{2}g^{ad}[f_{db}^{\ \
e}g_{ec}+f_{cb}^{\ \ e} g_{de}- f_{cd}^{\ \ e}g_{be}].$$ The $D$'s
can be related to $f$'s by a rescaling \cite{Review}: $$ D^{a}_{\
bc}=(\lambda^{a}\lambda^{b}/\lambda^{c})f^{a}_{\ bc},$$ where the
$\lambda$'s depend on the coset radii. Note that in general the
rescalings change the antisymmetry properties of $f$'s, while
 in the case of equal radii $D^{a}_{\ bc}=\frac{1}{2}f^{a}_{\ bc}$.
Note also that the connection-form $\omega^{a}_{\ b}$ is
$S$-invariant. This means that parallel transport commutes with
the $S$ action \cite{Castellani}. Then the most general form of an
$S$-invariant connection on $S/R$ would be
\begin{equation}
\omega^{a}_{\ b} = f_{\ \ ib}^{a}e^{i}+J_{\ \ cb}^{a}e^{c},
\end{equation}
with $J$ an $R$-invariant tensor, i.e. $$ \delta J_{cb}^{\ \ a}
=-f_{ic}^{\ \ d}J_{db}^{\ \ a}+ f_{id}^{\ \ a}J_{cb}^{\ \
d}-f_{ib}^{\ \ d}J_{cd}^{\ \ a}=0. $$ This condition is satisfied
by the $D$'s as can be proven using the Jacobi identity.

 In the case of non-vanishing torsion we have
\begin{equation}
T^{a} = de^{a} + \theta^{a}_{\ b} \wedge e^{b},
\end{equation}
where $$\theta^{a}_{\ b}=\omega^{a}_{\ b}+\tau^{a}_{\ b},$$ with
\begin{equation}
\tau^{a}_{\ b} = - \frac{1}{2} \Sigma^{a}_{\ bc}e^{c},
\end{equation}
while the contorsion $ \Sigma^{a}_{\ \ bc} $ is given by
\begin{equation}
\Sigma^{a}_{\ \ bc} = T^{a}_{\ \ bc}+T_{bc}^{\ \ a}-T_{cb}^{\ \ a}
\end{equation}
in terms of the torsion components $ T^{a}_{\ \ bc} $. Therefore
in general the connection-form $ \theta^{a}_{\ b}$ is
\begin{equation}
\theta^{a}_{\ b} = -f^{a}_{\ ic}e^{i} -(D^{a}_{\
bc}+\frac{1}{2}\Sigma^{a}_{\ bc})e^{c}= -f^{a}_{\
ic}e^{i}-G^{a}_{\ bc}e^{c}.
\end{equation}
The natural choice of torsion which would generalize the case of
equal radii \cite{DLust,Gavrilik,Batakis}, $T^{a}_{\ bc}=\eta
f^{a}_{\ bc}$ would be $T^{a}_{\ bc}=2\tau D^{a}_{\ bc}$ except
that the $D$'s do not have the required symmetry properties.
Therefore we must define $\Sigma$ as a combination of $D$'s which
makes $\Sigma$ completely antisymmetric  and $S$-invariant
according to the definition given above. Thus we are led to the
definition
\begin{equation}
\Sigma_{abc} \equiv 2\tau(D_{abc}+D_{bca}-D_{cba}).
\end{equation}
 In this general case the Riemann curvature two-form is given by
\cite{Review}, \cite{Batakis}:
\begin{equation}
R^{a}_{\ b}=[-\frac{1}{2}f_{ib}^{\ a}f_{de}^{\ i}-
\frac{1}{2}G_{cb}^{\ a}f_{de}^{\ c}+ \frac{1}{2}(G_{dc}^{\
a}G_{eb}^{\ c}-G_{ec}^{\ a}G_{db}^{\ c})]e^{d} \wedge e^{e},
\end{equation}
whereas the Ricci tensor $R_{ab}=R^{d}_{\ adb}$ is
\begin{equation}
R_{ab}=G_{ba}^{\ c}G_{dc}^{\ d}-G_{bc }^{\ d}G_{da }^{\ c}-G_{ca
}^{\ d}f_{db }^{\ c}-f_{ia }^{\ d}f_{db }^{\ i}.
\end{equation}
By choosing the parameter $\tau$ to be equal to zero we can obtain
the { \it Riemannian connection} $\theta_{R \ \ b}^{\ a}$. We can
also define the { \it canonical connection} by adjusting the radii
and $\tau$ so that the connection form is $\theta_{C \ \ b}^{\ a}
= -f^{a}_{\ bi}e^{i}$, i.e. an $R$-gauge field \cite{DLust}. The
adjustments should be such that $G_{abc}=0$. In the case of
$G_{2}/SU(3)$ where the metric is $g_{ab}=a\delta_{ab}$, we have $
G_{abc}=\frac{1}{2}a(1+3\tau)f_{abc}$ and in turn $\tau=
-\frac{1}{3}$. In the case of $Sp(4)/(SU(2) \times
U(1))_{non-max.}$, where the metric is $g_{ab}=diag(a,a,b,b,a,a)$,
we have to set $a=b$ and then $\tau=- \frac{1}{3}$ to obtain the
canonical connection. Similarly in the case of $SU(3)/(U(1) \times
U(1))$, where the metric is $g_{ab}=diag(a,a,b,b,c,c)$, we should
set $a=b=c$ and take $\tau=- \frac{1}{3}$. By analogous
adjustments we can set the Ricci tensor equal to zero
\cite{DLust}, thus defining a {\it Ricci flattening connection}.
\subsection{Reduction of a $D$-dimensional Yang-Mills-Dirac
Lagrangian.}
 The group $S$ acts as a symmetry group on the extra
coordinates. The CSDR scheme demands that an $S$-transformation
of the extra $d$ coordinates is a gauge transformation of the
fields that are defined on $M^{4}\times S/R$,  thus a gauge
invariant Lagrangian written on this space is independent of the
extra coordinates.

To see this in detail we consider a $D$-dimensional
Yang-Mills-Dirac theory with gauge group $G$ defined on a
manifold $M^{D}$ which as stated will be compactified to
$M^{4}\times S/R$, $D=4+d$, $d=dimS-dimR$:
\begin{equation}
A=\int d^{4}xd^{d}y\sqrt{-g}\Bigl[-\frac{1}{4}
Tr\left(F_{MN}F_{K\Lambda}\right)g^{MK}g^{N\Lambda}
+\frac{i}{2}\overline{\psi}\Gamma^{M}D_{M}\psi\Bigr] ,
\end{equation}
where
\begin{equation}
D_{M}= \partial_{M}-\theta_{M}-A_{M},
\end{equation}
with
\begin{equation}
\theta_{M}=\frac{1}{2}\theta_{MN\Lambda}\Sigma^{N\Lambda}
\end{equation}
the spin connection of $M^{D}$, and
\begin{equation}
F_{MN}
=\partial_{M}A_{N}-\partial_{N}A_{M}-\left[A_{M},A_{N}\right],
\end{equation}
where $M$, $N$ run over the $D$-dimensional space. The fields
$A_{M}$ and $\psi$ are, as explained, symmetric in the sense that
any transformation under symmetries of $S/R$  is compensated by
gauge transformations. The fermion fields can be in any
representation $F$ of $G$ unless a further symmetry such as
supersymmetry is required. So let $\xi_{A}^{\alpha}$, $A
=1,\ldots,dimS$, be the Killing vectors which generate the
symmetries of $S/R$ and $W_{A}$ the compensating gauge
transformation associated with $\xi_{A}$. Define next the
infinitesimal coordinate transformation as $\delta_{A} \equiv
L_{\xi_{A}}$, the Lie derivative with respect to $\xi$, then we
have for the scalar,vector and spinor fields,
\begin{eqnarray}
\delta_{A}\phi&=&\xi_{A}^{\alpha}\partial_{\alpha}\phi=D(W_{A})\phi,
\nonumber \\
\delta_{A}A_{\alpha}&=&\xi_{A}^{\beta}\partial_{\beta}A_{\alpha}+\partial_{
\alpha}
\xi_{A}^{\beta}A_{\beta}=\partial_{\alpha}W_{A}-[W_{A},A_{\alpha}],
\label{pan:21}\\
\delta_{A}\psi&=&\xi_{A}^{\alpha}\psi-\frac{1}{2}G_{Abc}\Sigma^{bc}\psi=
D(W_{A})\psi. \nonumber
\end{eqnarray}
$W_{A}$ depend only on internal coordinates $y$ and $D(W_{A})$
represents a gauge transformation in the appropriate
representation of the fields. $G_{Abc}$ represents a tangent space
rotation of the spinor fields. The variations $\delta_{A}$
satisfy, $[\delta_{A},\delta_{B}]=f_{AB}^{\\C}\delta_{C}$ and
lead to the following consistency relation for $W_{A}$'s,
\begin{equation}
\xi_{A}^{\alpha}\partial_{\alpha}W_{B}-\xi_{B}^{\alpha}\partial_{\alpha}
W_{A}-\left[W_{A},W_{B}\right]=f_{AB}^{\ \ C}W_{C}.
\end{equation}
 Furthermore the W's themselves transform under a gauge
transformation \cite{Review} as,
\begin{equation}\label{pan:23}
\widetilde{W}_{A} = gW_{A}g^{-1}+(\delta_{A}g)g^{-1}.
\end{equation}
Using \eqref{pan:23} and the fact that the Lagrangian is independent of
$y$ we can do all calculations at $y=0$ and choose a gauge where
$W_{a}=0$.

The detailed analysis of the constraints \eqref{pan:21} given in
refs.\cite{Manton,Review} provides us with the four-dimensional
unconstrained fields as well as with the gauge invariance that
remains in the theory after dimensional reduction. Here we give
the results. The components $A_{\mu}(x,y)$ of the initial gauge
field $A_{M}(x,y)$ become, after dimensional reduction, the
four-dimensional gauge fields and furthermore they are independent
of $y$. In addition one can find that they have to commute with
the elements of the $R_{G}$ subgroup of $G$. Thus the
four-dimensional gauge group $H$ is the centralizer of $R$ in $G$,
$H=C_{G}(R_{G})$. Similarly, the $A_{\alpha}(x,y)$ components of
$A_{M}(x,y)$ denoted by $\phi_{\alpha}(x,y)$ from now on, become
scalars at four dimensions. These fields transform under $R$ as a
vector $v$, i.e.
\begin{eqnarray}
S &\supset& R \nonumber \\
adjS &=& adjR+v.
\end{eqnarray}
Moreover $\phi_{\alpha}(x,y)$ act as an intertwining operator
connecting induced representations of $R$ acting on $G$ and $S/R$.
This implies, exploiting Schur's lemma, that the transformation
properties of the fields $\phi_{\alpha}(x,y)$ under $H$ can be
found if we express the adjoint representation of $G$ in terms of
$R_{G} \times H$ :
\begin{eqnarray}
G &\supset& R_{G} \times H \nonumber \\
 adjG &=&(adjR,1)+(1,adjH)+\sum(r_{i},h_{i}).
\end{eqnarray}
Then if $v=\sum s_{i}$, where each $s_{i}$ is an irreducible
representation of $R$, there survives an $h_{i}$ multiplet for
every pair $(r_{i},s_{i})$, where $r_{i}$ and $s_{i}$ are
identical irreducible representations of $R$.

Turning next to the fermion fields
\cite{Review,Slansky,Chapline,Palla} similarly to scalars, they
act as intertwining operators between induced representations
acting on $G$ and the tangent space of $S/R$, $SO(d)$. Proceeding
along similar lines as in the case of scalars to obtain the
representation of $H$ under which the four-dimensional fermions
transform, we have to decompose the representation $F$ of the
initial gauge group in which the fermions are assigned under
$R_{G} \times H$, i.e.
\begin{equation}\label{pan:26}
F= \sum (t_{i},h_{i}),
\end{equation}
and the spinor of $SO(d)$ under $R$
\begin{equation}\label{pan:27}
\sigma_{d} = \sum \sigma_{j}.
\end{equation}
Then for each pair $t_{i}$ and $\sigma_{i}$, where $t_{i}$ and
$\sigma_{i}$ are identical irreducible representations there is an
$h_{i}$ multiplet of spinor fields in the four-dimensional theory.
In order however  to obtain chiral fermions in the effective
theory we have to impose further requirements. We first impose the
Weyl condition in $D$ dimensions. In $D = 4n+2$ dimensions which
is the case at hand, the decomposition of the left handed, say
spinor under $SU(2) \times SU(2) \times SO(d)$ is
\begin{equation}
\sigma _{D} = (2,1,\sigma_{d}) + (1,2,\overline{\sigma}_{d}).
\end{equation}
So we have in this case the decompositions
\begin{equation}
\sigma_{d} = \sum \sigma_{k},~\overline{\sigma}_{d}= \sum
\overline{\sigma}_{k}.
\end{equation}
Let us start from a vector-like representation $F$ for the
fermions. In this case each term $(t_{i},h_{i})$ in \eqref{pan:26} will
be either self-conjugate or it will have a partner $(
\overline{t}_{i},\overline{h}_{i} )$. According to the rule
described in eqs.\eqref{pan:26}, \eqref{pan:27} and considering $\sigma_{d}$ we
will
have in four dimensions left-handed fermions transforming as $
f_{L} = \sum h^{L}_{k}$. It is important to notice that since
$\sigma_{d}$ is non self-conjugate, $f_{L}$ is non self-conjugate
too. Similarly from $\overline{\sigma}_{d}$ we will obtain the
right handed representation $ f_{R}= \sum \overline{h}^{R}_{k}$
but as we have assumed that $F$ is vector-like,
$\overline{h}^{R}_{k}\sim h^{L}_{k}$. Therefore there will appear
two sets of Weyl fermions with the same quantum numbers under $H$.
This is already a chiral theory, but still one can go further and
try to impose the Majorana condition in order to eliminate the
doubling of the fermionic spectrum. We should remark now that if
we had started with $F$ complex, we should have again a chiral
theory since in this case $\overline{h}^{R}_{k}$ is different from
$h^{L}_{k}$  $(\sigma_{d}$ non self-conjugate). Nevertheless
starting with $F$ vector-like is much more appealing and will be
used in the following along with the Majorana condition. Majorana and Weyl
conditions are compatible in $D=4n+2$ dimensions. Then in our case
if we start with Weyl-Majorana spinors in $D=4n+2$ dimensions we
force $f_{R}$ to be the charge conjugate to $f_{L}$, thus arriving
in a theory with fermions only in $f_{L}$.

An important requirement is that the resulting four-dimensional theories should
be anoma\-ly free.
Starting with an anomaly free theory in higher dimensions,  Witten~\cite{Witten}
 has given the
condition to be fulfilled in order to obtain anomaly free four-dimensional
theories. The condition restricts
the allowed embeddings of $R$ into $G$ by relating them with the embedding of
$R$ into $SO(6)$, the tangent
space of the six-dimensional cosets we consider~\cite{Review,Pilch}. To be more
specific if
$L_{a}$ are the generators of $R$ into $G$ and $T_{a}$ are the generators of $R$
into $SO(6)$ the condition
reads
\begin{equation}\label{pan:30}
Tr(L_{a} L_{b})=30\,Tr(T_{a}T_{b})\,.
\end{equation}
According to ref.~\cite{Pilch} the anomaly cancellation
condition~(\ref{pan:30}) is automatically
satisfied for the choice of embedding
\begin{equation}
E_{8}\supset SO(6) \supset R\,,\label{RinSO6inGembedding}
\end{equation}
which we adopt here. Furthermore concerning the abelian group factors of the
four-dimensional gauge theory,
we note that the corresponding gauge bosons surviving in four dimensions become
massive at the
compactification scale~\cite{Witten,Green} and therefore, they do not contribute
in the
anomalies;  they correspond only to global symmetries.

\subsection{The Four-Dimensional Theory.}\label{pan:s2.3}
Next let us obtain the four-dimensional effective action. Assuming
that the metric is block diagonal, taking into account all the
constraints and integrating out the extra coordinates we obtain in
four dimensions the following Lagrangian :
\begin{equation}\label{pan:31}
A=C \int d^{4}x \biggl( -\frac{1}{4} F^{t}_{\mu
\nu}{F^{t}}^{\mu\nu}+\frac{1}{2}(D_{\mu}\phi_{\alpha})^{t}
(D^{\mu}\phi^{\alpha})^{t}
+V(\phi)+\frac{i}{2}\overline{\psi}\Gamma^{\mu}D_{\mu}\psi-\frac{i}{2}
\overline{\psi}\Gamma^{a}D_{a}\psi\biggr),
\end{equation}
where $D_{\mu} = \partial_{\mu} - A_{\mu}$ and $D_{a}=
\partial_{a}- \theta_{a}-\phi_{a}$ with  $\theta_{a}=
\frac{1}{2}\theta_{abc}\Sigma^{bc}$ the connection of the coset
space, while $C$ is the volume of the coset space. The potential
$V(\phi)$ is given by:
\begin{equation}\label{pan:32}
V(\phi) = - \frac{1}{4} g^{ac}g^{bd}Tr( f _{ab}^{C}\phi_{C} -
[\phi_{a},\phi_{b}] ) (f_{cd}^{D}\phi_{D} - [\phi_{c},\phi_{d}] )
,
\end{equation}
where, $A=1,\ldots,dimS$ and $f$ ' s are the structure constants
appearing in the commutators of the generators of the Lie algebra
of S. The expression \eqref{pan:32} for $V(\phi)$ is only formal because
$\phi_{a}$ must satisfy the constraints coming from eq.\eqref{pan:21},
\begin{equation}\label{pan:33}
f_{ai}^{D}\phi_{D} - [\phi_{a},\phi_{i}] = 0,
\end{equation}
where the $\phi_{i}$ generate $R_{G}$. These constraints imply
that some components $\phi_{a}$'s are zero, some are constants and
the rest can be identified with the genuine Higgs fields. When
$V(\phi)$ is expressed in terms of the unconstrained independent
Higgs fields, it remains a quartic polynomial which is invariant
under gauge transformations of the final gauge group $H$, and its
minimum determines the vacuum expectation values of the Higgs
fields \cite{Vinet,Harnad,Farakos}. The minimization of the
potential is in general a difficult problem. If however $S$ has an
isomorphic image $S_{G}$ in $G$ which contains $R_{G}$ in a
consistent way then it is possible to allow the $\phi_{a}$ to
become generators of $S_{G}$. That is $\overline{\phi}_{a} =
<\phi^{i}>Q_{ai} = Q_{a}$ with $<\phi^{i}>Q_{ai}$ suitable
combinations of $G$ generators, $Q_{a}$ a generator of $S_{G}$
and $a$ is also a coset-space index. Then
\begin{eqnarray*}
\overline{F}_{ab}&=&f_{ab}^{\ \ i}Q_{i}+f_{ab}^{\ \
c}\overline{\phi}_{c}-[\overline{\phi}_{a},\overline{\phi}_{b}]\\
&=& f_{ab}^{\ \ i}Q_{i}+ f_{ab}^{\ \ c}Q_{c}- [Q_{a},Q_{b}] = 0
\end{eqnarray*}
because of the commutation relations of $S$. Thus we  have proven
that $V(\phi=\overline{\phi})=0$ which furthermore is the minimum,
because $V$ is positive definite. Furthermore, the
four-dimensional gauge group $H$ breaks further by these non-zero
vacuum expectation values of the Higgs fields to the centralizer
$K$ of the image of $S$ in $G$, i.e. $K=C_{G}(S)$
\cite{Review,Vinet,Harnad,Farakos}. This can been seen if we
examine a gauge transformation of $\phi_{a}$ by an element $h$ of
$H$. Then we have $$ \phi_{a} \rightarrow h\phi_{a}h^{-1}, h \in
H $$ We note that the v.e.v. of the Higgs fields is gauge
invariant for the set of $h$'s that commute with $S$. That is $h$
belongs to a subgroup $K$ of $H$ which is the centralizer of
$S_{G}$ in $G$.

In the fermion part of the Lagrangian the first term is just the
kinetic term of fermions, while the second is the Yukawa term
\cite{Kapetanakis}. Note that since $\psi$ is a Majorana-Weyl
spinor in ten dimensions the representation in which the fermions
are assigned under the gauge group must be real. The last term in
\eqref{pan:31} can be written as
\begin{equation}\label{pan:38}
L_{Y}= -\frac{i}{2}\overline{\psi}\Gamma^{a}(\partial_{a}-
\frac{1}{2}f_{ibc}e^{i}_{\Gamma}e^{\Gamma}_{a}\Sigma^{bc}-
\frac{1}{2}G_{abc}\Sigma^{bc}- \phi_{a}) \psi \nonumber \\
=\frac{i}{2}\overline{\psi}\Gamma^{a}\nabla_{a}\psi+
\overline{\psi}V\psi ,
\end{equation}
where
\begin{eqnarray}
\nabla_{a}& =& - \partial_{a} +
\frac{1}{2}f_{ibc}e^{i}_{\Gamma}e^{\Gamma}_{a}\Sigma^{bc} + \phi_{a},
\label{pan:39}\\
 V&=&\frac{i}{4}\Gamma^{a}G_{abc}\Sigma^{bc},\label{pan:40}
\end{eqnarray}
and we have used the full connection with torsion \cite{Batakis}
given by
\begin{equation}
\theta_{\ \ c b}^{a} = - f_{\ \
ib}^{a}e^{i}_{\alpha}e^{\alpha}_{c}-(D_{\ \ cb}^{a} +
\frac{1}{2}\Sigma_{\ \ cb}^{a}) = - f_{\ \
ib}^{a}e^{i}_{\alpha}e^{\alpha}_{c}- G_{\ \ cb}^{a}
\end{equation}
with
\begin{equation}
D_{\ \ cb}^{a} = g^{ad}\frac{1}{2}[f_{db}^{\ \ e}g_{ec} + f_{
cb}^{\ \ e}g_{de} - f_{cd}^{\ \ e}g_{be}]
\end{equation}
and
\begin{equation}
\Sigma_{abc}= 2\tau(D_{abc} +D_{bca} - D_{cba}).
\end{equation}
 We have already noticed that the CSDR constraints tell us that
$\partial_{a}\psi= 0$. Furthermore we can consider the Lagrangian
at the point $y=0$, due to its invariance under
$S$-transformations, and as we mentioned $e^{i}_{\Gamma}=0$ at
that point. Therefore \eqref{pan:39} becomes just $\nabla_{a}= \phi_{a}$
and the term $\frac{i}{2}\overline{\psi}\Gamma^{a}\nabla_{a}\psi $
in \eqref{pan:38} is exactly the Yukawa term.

Let us examine now the last term appearing in \eqref{pan:38}. One can
show easily that the operator $V$ anticommutes with the
six-dimensional helicity operator \cite{Review}. Furthermore one
can show that $V$ commutes with the $T_{i}=
-\frac{1}{2}f_{ibc}\Sigma^{bc}$ ($T_{i}$ close the $R$-subalgebra
of $SO(6)$). In turn we can draw the conclusion, exploiting
Schur's lemma, that the non-vanishing elements of $V$ are only
those which appear in the decomposition of both $SO(6)$ irreps $4$
and $\overline{4}$, e.g. the singlets. Since this term is of pure
geometric nature, we reach the conclusion that the singlets in $4$
and $\overline{4}$ will acquire large geometrical masses, a fact
that has serious phenomenological implications. In supersymmetric
theories defined in higher dimensions, it means that the gauginos
obtained in four dimensions after dimensional reduction receive
masses comparable to the compactification scale. However as we
shall see in the next section this result changes in presence of
torsion. We note that for symmetric coset spaces the $V$ operator
is absent because $f_{ab}^{c}$ are vanishing by definition in that
case.

\section{Dimensional Reduction of $E_8$ over $SU(3)/U(1)\times U(1)$ and
soft supersymmetry breaking\label{pan:s3.2}}

In this model we consider the coset space $B=SU(3)/U(1) \times U(1)$ on which we reduce the 
ten-dimensional theory.
To determine the four-dimensional gauge group, the embedding of $R = U(1) \times U(1)$ in $E_8$ is suggested by the decomposition
\begin{equation}
E_8 \supset E_6\times SU(3) \supset E_6 \times U(1)_A \times U(1)_B.
\end{equation}
Then, the surviving gauge group in four dimensions is 
$$ H=C_{E_{8}}(U(1) \times U(1)) =   E_{6}\times U(1)_A \times U(1)_B. $$
The $248$ of $E_{8}$ decomposes under $ E_{6}\times  U(1)_A \times U(1)_B$ in
the
following way:
\begin{eqnarray}
 248 = 1_{(0,0)}+1_{(0,0)}+1_{(3,\frac{1}{2})}+1_{(-3,\frac{1}{2})}+\nonumber\\
1_{(0,-1)}+1_{(0,1)}+1_{(-3,-\frac{1}{2})}+1_{(3,-\frac{1}{2})}+\nonumber\\
78_{(0,0)}+27_{(3,\frac{1}{2})}+27_{(-3,\frac{1}{2})}+27_{(0,-1)}+\nonumber\\
\overline{27}_{(-3,-\frac{1}{2})}+\overline{27}_{(3,-\frac{1}{2})}
+\overline{27}_{(0,1)}.\label{pan:78}
\end{eqnarray}
The $R=U(1) \times U(1)$ decomposition of the vector and spinor representations of $SO(6)$
(see table \ref{table4}, Appendix \ref{tables}) is
$${\boldsymbol{6_\upsilon}}=(3,\frac{1}{2})+(-3,\frac{1}{2})
+(0,-1)+(-3,-\frac{1}{2})+(3,-\frac{1}{2})+(0,1)$$ and
$${\boldsymbol{4_s}}=(0,0)+(3,\frac{1}{2})+(-3,\frac{1}{2}) +(0,-1)$$ 
respectively.
Thus applying the CSDR rules we find that the surviving fields in
four dimensions are three ${\cal N}=1$ vector multiplets
$V^{\alpha},V_{(1)},V_{(2)}$, (where $\alpha$ is an $E_{6}$, $78$
index and the other two refer to the two $U(1)'s$) containing the
gauge fields of $ E_{6}\times U(1)_A \times U(1)_B$. The matter
content consists of three ${\cal N}=1$ chiral multiplets ($A^{i}$,
$B^{i}$, $C^{i}$) with $i$ an $E_{6}$, $27$ index and three ${\cal
N}=1$ chiral multiplets ($A$, $B$, $C$) which are $E_{6}$ singlets
and carry only $U(1)_A \times U(1)_B$ charges.

To determine the potential we examine further the decomposition
of the adjoint of the specific $S=SU(3)$ under $R=U(1) \times
U(1)$, i.e.
$$SU(3) \supset U(1) \times U(1) $$
\begin{eqnarray}
 8 = (0,0)+(0,0)+6_{\upsilon}. \label{pan:79}
\end{eqnarray}
Then according to the decomposition \eqref{pan:79} the generators of $SU(3)$
can be grouped as
\begin{equation}\label{pan:80}
Q_{SU(3)} = \{Q_{0},Q'_{0},Q_{1},Q_{2},Q_{3},Q^{1},Q^{2},Q^{3} \}.
\end{equation}
The non trivial commutator relations of $SU(3)$ generators \eqref{pan:80}
are given in the table \ref{table1} given in the Appendix \ref{tables}.
The decomposition \eqref{pan:80}
suggests the following change in  the notation of the scalar
fields,
\begin{equation}\label{pan:81}
(\phi_{I}, I=1,\ldots,8) \longrightarrow ( \phi_{0}, \phi'_{0},
\phi_{1}, \phi^{1}, \phi_{2}, \phi^{2}, \phi_{3}, \phi^{3}).
\end{equation}

The potential of any theory reduced over $SU(3)/U(1) \times U(1))$
is given in terms of the redefined fields in \eqref{pan:81} by
\begin{eqnarray}
\lefteqn{V(\phi)=(3\Lambda^{2}+\Lambda'^{2})\biggl(\frac{1}{R_{1}^{4}}+\frac{1}{R_{2}^{4}}\biggr)
+\frac{4\Lambda'^{2}}{R_{3}^{2}}}\nonumber\\
&&+\frac{2}{R_{2}^{2}R_{3}^{2}}Tr(\phi_{1}\phi^{1})+
\frac{2}{R_{1}^{2}R_{3}^{2}}Tr(\phi_{2}\phi^{2})
+\frac{2}{R_{1}^{2}R_{2}^{2}}Tr(\phi_{3}\phi^{3})\nonumber\\
&&+\frac{\sqrt{3}\Lambda}{R_{1}^{4}}Tr(Q_{0}[\phi_{1},\phi^{1}])
-\frac{\sqrt{3}\Lambda}{R_{2}^{4}}Tr(Q_{0}[\phi_{2},\phi^{2}])
-\frac{\sqrt{3}\Lambda}{R_{3}^{4}}Tr(Q_{0}[\phi_{3},\phi^{3}])\nonumber\\
&&+\frac{\Lambda'}{R_{1}^{4}}Tr(Q'_{0}[\phi_{1},\phi^{1}])
+\frac{\Lambda'}{R_{2}^{4}}Tr(Q'_{0}[\phi_{2},\phi^{2}])
-\frac{2\Lambda'}{R_{3}^{4}}Tr(Q'_{0}[\phi_{3},\phi^{3}])\nonumber\\
&&+\biggl[\frac{2\sqrt{2}}{R_{1}^{2}R_{2}^{2}}Tr(\phi_{3}[\phi_{1},\phi_{2}])
+\frac{2\sqrt{2}}{R_{1}^{2}R_{3}^{3}}Tr(\phi_{2}[\phi_{3},\phi_{1}])
+\frac{2\sqrt{2}}{R_{2}^{2}R_{3}^{2}}Tr(\phi_{1}[\phi_{2},\phi_{3}])+ h.c\biggr]\nonumber\\
&&+\frac{1}{2}Tr \biggl(\frac{1}{R_{1}^{2}}([\phi_{1},\phi^{1}])+
\frac{1}{R_{2}^{2}}([\phi_{2},\phi^{2}])
+\frac{1}{R_{3}^{2}}([\phi_{3},\phi^{3}])\biggr)^{2}\nonumber\\
&&-\frac{1}{R_{1}^{2}R_{2}^{2}}Tr([\phi_{1},\phi_{2}][\phi^{1},\phi^{2}])
-\frac{1}{R_{1}^{2}R_{3}^{2}}Tr([\phi_{1},\phi_{3}][\phi^{1},\phi^{3}])\nonumber\\
&&-\frac{1}{R_{2}^{2}R_{3}^{2}}Tr([\phi_{2},\phi_{3}][\phi^{2},\phi^{3}]),
\label{pan:82}
\end{eqnarray}
where $R_{1},R_{2},R_{3}$ are the coset space radii\footnote{To
bring the potential into this form we have used (A.22) of
ref.\cite{Review} and relations (7),(8) of ref.\cite{Witten2}.}.
In terms of the radii the real metric\footnote{The complex metric
that was used is
$g^{1\overline{1}}=\frac{1}{R_{1}^{2}},g^{2\overline{2}}=\frac{1}{R_{2}^{2}},
g^{3\overline{3}}=\frac{1}{R_{3}^{2}}$.} of the coset is
\begin{equation}
g_{ab}=diag(R_{1}^{2},R_{1}^{2},R_{2}^{2},R_{2}^{2},R_{3}^{2},R_{3}^{2}).
\end{equation}

Next we examine the commutation relations of $E_{8}$ under the
decomposition \eqref{pan:78}. Under this decomposition the generators of
$E_{8}$ can be grouped as
\begin{eqnarray}
Q_{E_{8}}=\{Q_{0},Q'_{0},Q_{1},Q_{2},Q_{3},Q^{1},Q^{2},Q^{3},Q^{\alpha},\nonumber\\
Q_{1i},Q_{2i},Q_{3i},Q^{1i},Q^{2i},Q^{3i} \},\label{pan:84}
\end{eqnarray}
where, $ \alpha=1,\ldots,78 $ and $ i=1,\ldots,27 $. The
non-trivial commutation relations of the $E_{8}$ generators \eqref{pan:84}
are given in Appendix \ref{tables} in the table \ref{table2}. 
\\ Now the
constraints \eqref{pan:33} for the redefined fields in \eqref{pan:81} are,
\begin{eqnarray}
\left[\phi_{1},\phi_{0}\right]=\sqrt{3}\phi_{1}&,&
\left[\phi_{1},\phi_{0}'\right]=\phi_{1}, \nonumber \\
\left[\phi_{2},\phi_{0}\right]=-\sqrt{3}\phi_{2}&,&
\left[\phi_{2},\phi_{0}'\right]=\phi_{2}, \nonumber \\
\left[\phi_{3},\phi_{0}\right]=0&,&
\left[\phi_{3},\phi_{0}'\right]=-2\phi_{3}.\label{pan:85}
\end{eqnarray}
The solutions of the constraints \eqref{pan:85} in terms of the genuine
Higgs fields and of the $E_{8}$ generators \eqref{pan:84} corresponding to
the embedding \eqref{pan:78} of $R=U(1) \times U(1)$ in  the $E_{8}$ are,
$\phi_{0}=\Lambda Q_{0}$ and $\phi_{0}'=\Lambda Q_{0}'$,with
$\Lambda=\Lambda'=\frac{1}{\sqrt{10}}$, and
\begin{eqnarray}
\phi_{1} &=& R_{1} \alpha^{i} Q_{1i}+R_{1} \alpha Q_{1},
\nonumber\\ \phi_{2} &=& R_{2} \beta^{i} Q_{2i}+ R_{2} \beta
Q_{2}, \nonumber\\ \phi_{3} &=& R_{3} \gamma^{i} Q_{3i}+ R_{3}
\gamma Q_{3}, \label{pan:86}
\end{eqnarray}
where the unconstrained  scalar fields transform under $  E_{6}\times U(1)_A
\times U(1)_B$ as
\begin{eqnarray}
\alpha_{i} \sim 27_{(3,\frac{1}{2})}&,&\alpha \sim
1_{(3,\frac{1}{2})},\nonumber\\ \beta_{i} \sim
27_{(-3,\frac{1}{2})}&,&\beta \sim
1_{(-3,\frac{1}{2})},\nonumber\\ \gamma_{i} \sim
27_{(0,-1)}&,&\gamma \sim 1_{(0,-1)}.\label{pan:87}
\end{eqnarray}
The potential \eqref{pan:82} becomes
\begin{eqnarray}
V(\alpha^{i},\alpha,\beta^{i},\beta,\gamma^{i},\gamma)= const. +
\biggl( \frac{4R_{1}^{2}}{R_{2}^{2}R_{3}^{2}}-\frac{8}{R_{1}^{2}}
\biggr)\alpha^{i}\alpha_{i} +\biggl(
\frac{4R_{1}^{2}}{R_{2}^{2}R_{3}^{2}}-\frac{8}{R_{1}^{2}}
\biggr)\overline{\alpha}\alpha \nonumber \\
+\biggl(\frac{4R_{2}^{2}}{R_{1}^{2}R_{3}^{2}}-\frac{8}{R_{2}^{2}}\biggr)
\beta^{i}\beta_{i}
+\biggl(\frac{4R_{2}^{2}}{R_{1}^{2}R_{3}^{2}}-\frac{8}{R_{2}^{2}}\biggr)
\overline{\beta}\beta \nonumber \\
+\biggl(\frac{4R_{3}^{2}}{R_{1}^{2}R_{2}^{2}}
-\frac{8}{R_{3}^{2}}\biggr)\gamma^{i}\gamma_{i}
+\biggl(\frac{4R_{3}^{2}}{R_{1}^{2}R_{2}^{2}}
-\frac{8}{R_{3}^{2}}\biggr)\overline{\gamma}\gamma \nonumber\\
+\biggl[\sqrt{2}80\biggl(\frac{R_{1}}{R_{2}R_{3}}+\frac{R_{2}}{R_{1}
R_{3}}+\frac{R_{3}}{R_{2}R_{1}}\biggr)d_{ijk}\alpha^{i}\beta^{j}\gamma^{k}\nonumber\\
+\sqrt{2}80\biggl(\frac{R_{1}}{R_{2}R_{3}}+\frac{R_{2}}{R_{1}
R_{3}}+\frac{R_{3}}{R_{2}R_{1}}\biggr)\alpha\beta\gamma+
h.c\biggr]\nonumber\\
+\frac{1}{6}\biggl(\alpha^{i}(G^{\alpha})_{i}^{j}\alpha_{j}
+\beta^{i}(G^{\alpha})_{i}^{j}\beta_{j}
+\gamma^{i}(G^{\alpha})_{i}^{j}\gamma_{j}\biggr)^{2}\nonumber\\
+\frac{10}{6}\biggl(\alpha^{i}(3\delta_{i}^{j})\alpha_{j} +
\overline{\alpha}(3)\alpha + \beta^{i}(-3\delta_{i}^{j})\beta_{j}
+ \overline{\beta}(-3)\beta \biggr)^{2}\nonumber \\
+\frac{40}{6}\biggl(\alpha^{i}(\frac{1}{2}\delta_{i}^{j})\alpha_{j}
+ \overline{\alpha}(\frac{1}{2})\alpha +
\beta^{i}(\frac{1}{2}\delta^{j}_{i})\beta_{j} +
\overline{\beta}(\frac{1}{2})\beta +
\gamma^{i}(-1\delta_{i}^{j})\gamma_{j} +
\overline{\gamma}(-1)\gamma \biggr)^{2}\nonumber \\
+40\alpha^{i}\beta^{j}d_{ijk}d^{klm}\alpha_{l}\beta_{m}
+40\beta^{i}\gamma^{j}d_{ijk}d^{klm}\beta_{l}\gamma_{m}
+40\alpha^{i}\gamma^{j}d_{ijk}d^{klm}\alpha_{l}\gamma_{m}\nonumber\\
+40(\overline{\alpha}\overline{\beta})(\alpha\beta) +
40(\overline{\beta}\overline{\gamma})(\beta\gamma) +
40(\overline{\gamma}\overline{\alpha})(\gamma\alpha).\label{pan:88}
\end{eqnarray}
From the potential \eqref{pan:88} we read the $F$-, $D$- and scalar soft
terms. The $F$-terms are obtained from the superpotential
\begin{equation}\label{pan:89}
{ \cal W }(A^{i},B^{j},C^{k},A,B,C)
=\sqrt{40}d_{ijk}A^{i}B^{j}C^{k} + \sqrt{40}ABC. 
\end{equation}
The $D$-terms have the structure
\begin{equation}
\frac{1}{2}D^{\alpha}D^{\alpha}+\frac{1}{2}D_{1}D_{1}+\frac{1}{2}D_{2}D_{2},
\end{equation}
where $$D^{\alpha}= \frac{1}{\sqrt{3}}
\biggl(\alpha^{i}(G^{\alpha})_{i}^{j}\alpha_{j}
+\beta^{i}(G^{\alpha})_{i}^{j}\beta_{j}
+\gamma^{i}(G^{\alpha})_{i}^{j}\gamma_{j}\biggr),$$ $$D_{1}=
\sqrt{ \frac{10}{3} }\biggl(\alpha^{i}(3\delta_{i}^{j})\alpha_{j}
+ \overline{\alpha}(3)\alpha +
\beta^{i}(-3\delta_{i}^{j})\beta_{j} + \overline{\beta}(-3)\beta
\biggr)$$ and $$D_{2} = \sqrt{ \frac{40}{3}
}\biggl(\alpha^{i}(\frac{1}{2}\delta_{i}^{j})\alpha_{j} +
\overline{\alpha}(\frac{1}{2})\alpha +
\beta^{i}(\frac{1}{2}\delta^{j}_{i})\beta_{j} +
\overline{\beta}(\frac{1}{2})\beta +
\gamma^{i}(-1\delta_{i}^{j})\gamma_{j} +
\overline{\gamma}(-1)\gamma \biggr),$$ which correspond to the
$E_{6} \times U(1)_A \times U(1)_B$ structure of the gauge
group. The rest terms are the trilinear and mass terms which break
supersymmetry softly and they form the scalar SSB part of the
Lagrangian,
\begin{eqnarray}
\lefteqn{{\cal L}_{scalarSSB}=  \biggl(
\frac{4R_{1}^{2}}{R_{2}^{2}R_{3}^{2}}-\frac{8}{R_{1}^{2}}
\biggr)\alpha^{i}\alpha_{i} +\biggl(
\frac{4R_{1}^{2}}{R_{2}^{2}R_{3}^{2}}-\frac{8}{R_{1}^{2}}
\biggr)\overline{\alpha}\alpha} \nonumber\\
& &
+\biggl(\frac{4R_{2}^{2}}{R_{1}^{2}R_{3}^{2}}-\frac{8}{R_{2}^{2}}\biggr)
\beta^{i}\beta_{i}
+\biggl(\frac{4R_{2}^{2}}{R_{1}^{2}R_{3}^{2}}-\frac{8}{R_{2}^{2}}\biggr)
\overline{\beta}\beta
+\biggl(\frac{4R_{3}^{2}}{R_{1}^{2}R_{2}^{2}}
-\frac{8}{R_{3}^{2}}\biggr)\gamma^{i}\gamma_{i}
+\biggl(\frac{4R_{3}^{2}}{R_{1}^{2}R_{2}^{2}}
-\frac{8}{R_{3}^{2}}\biggr)\overline{\gamma}\gamma \nonumber\\
& &
+\biggl[\sqrt{2}80\biggl(\frac{R_{1}}{R_{2}R_{3}}+\frac{R_{2}}{R_{1}
R_{3}}+\frac{R_{3}}{R_{2}R_{1}}\biggr)d_{ijk}\alpha^{i}\beta^{j}\gamma^{k}
\nonumber \\
& &+\sqrt{2}80\biggl(\frac{R_{1}}{R_{2}R_{3}}+\frac{R_{2}}{R_{1}
R_{3}}+\frac{R_{3}}{R_{2}R_{1}}\biggr)\alpha\beta\gamma+
h.c\biggr]. \label{pan:91}
\end{eqnarray}

Note that the potential \eqref{pan:88} belongs to the case analyzed in
subsection~\ref{pan:s2.3} where $S$ has an image in $G$. Here $S=SU(3)$ has
an image in $G=E_{8}$ \cite{LZ} so we conclude that the minimum of
the potential is zero. Finally in order to determine the gaugino
mass, we calculate the V operator of the eq. \eqref{pan:40} in the case of
$SU(3)/U(1) \times
U(1)$ using Appendix \ref{appC} and
using the $\Gamma$-matrices given in the Appendix \ref{appendix}
we calculate
$\Sigma^{ab}=\frac{1}{4}[\Gamma^{a},\Gamma^{b}]$ and then
$G_{abc}\Gamma^{a}\Sigma^{bc}$.
The combination of all leads to the gaugino mass 
\begin{equation}\label{pan:92}
M=V=(1+3\tau)\frac{(R_{1}^{2}+R_{2}^{2}+R_{3}^{2})}{8\sqrt{R_{1}^{2}R_{2}^{2}R_{
3 }^{2}}}.
\end{equation}
Note again that the chosen embedding satisfies the condition \eqref{pan:30}
and the absence in the four-dimensional theory of any other term
that does not belong to the supersymmetric $E_{6} \times U(1)_A
\times U(1)_B$ gauge theory or to its SSB sector. The gaugino mass
\eqref{pan:92} has a contribution from the
torsion of the coset space. A final remark concerning the gaugino
mass is that the adjustment
required to obtain the {\it canonical connection} leads also to
vanishing gaugino masses. Contrary to the gaugino mass term the
soft scalar terms of the SSB does not receive contributions from the
torsion. This is due to the fact that gauge fields, contrary to
fermions, do not couple to torsion.

 Concluding the present section, we would like to note that the fact that, 
starting with a ${\cal
N} = 1$ supersymmetric theory in ten dimensions, the CSDR leads to
the field content of an ${\cal N} = 1$ supersymmetric theory in
the case that the six-dimensional coset spaces used are
non-symmetric, can been seen by inspecting the table \ref{table4}. More
specifically, one notices in table \ref{table4} that when the coset spaces
are non-symmetric the decompositions of the spinor $4$ and
antispinor $\overline{4}$ of $SO(6)$ under $R$ contain a singlet,
i.e. have the form $1+r$ and $1+\overline{r}$, respectively, where
$r$ is possibly reducible. The singlet under $R$ provides the
four-dimensional theory with fermions transforming according to
the adjoint as was emphasized in subsection \ref{pan:s2.3} and correspond
to gauginos, which obtain geometrical and torsion mass
contributions as we have seen in  the present
section \ref{pan:s3.2}. Next turning the decomposition of the vector $6$
of $SO(6)$ under $R$ in the non-symmetric cases, we recall that
the vector can be constructed from the tensor product $4 \times 4$
and therefore has the form $r+\overline{r}$. Then the CSDR
constraints tell us that the four-dimensional theory will contain
the same representations of fermions and scalars since both come
from the adjoint representation of the gauge group $G$ and they
have to satisfy the same matching conditions under $R$. Therefore
the field content of the four-dimensional theory is, as expected,
${\cal N} =1$ supersymmetric. To find out that furthermore the
${\cal N} =1$ supersymmetry is softly broken, requires the lengthy
and detailed analysis that was done above.

\section{Wilson flux breaking}

Clearly, we need to further reduce the gauge symmetry. We will employ the Wilson
flux breaking mechanism \cite{Kozimirov,Zoupanos,Hosotani}.
Let us briefly recall the Wilson flux mechanism
 for breaking spontaneously a gauge theory. 

\subsection{The Wilson flux mechanism}

Instead of considering a gauge theory on $M^{4}\times B_{0}$, with $B_{0}$ a simply connected manifold, and in our case a
coset space $B_{0}=S/R$, we consider a gauge theory on $M^{4}\times B$, with $B=B_{0}/F^{S/R}$ and $F^{S/R}$ a
freely acting discrete symmetry of $B_{0}$. It turns out that $B$ becomes
multiply  connected, which means that there will be contours not contractible to a point due to holes in the
manifold. For each element $g\in F^{S/R}$, we  pick up an element $U_{g}$ in $H$, i.e. in the four-dimensional
gauge group of the reduced theory, which can be represented as the Wilson loop
(WL)
\begin{equation}
U_{g}=\mathcal{P}exp\left(-i~\int_{\gamma_g}T^{a}A_{M}^{a}(x)dx^{M}\right)\,,
\end{equation}
where $A_{M}^{a}(x)$ are vacuum $H$ fields with group generators $T^{a}$, $\gamma_g$ is a contour representing
the abstract element $g$ of $F^{S/R}$, and $\mathcal{P}$ denotes the path ordering.
Now if $\gamma_g$ is chosen not to be contractible to a point, then $U_g\neq 1$ although the vacuum field 
strength vanishes everywhere. In this  way an homomorphism of $F^{S/R}$ into $H$ is induced with image
$T^{H}$, which is the subgroup of $H$ generated by $\{U_g\}$. A field $f(x)$ on  $B_0$ is obviously equivalent
to another field on $B_0$ which obeys $f(g(x))=f(x)$ for every $g\in F^{S/R}$. However in the presence of the
gauge group $H$ this statement can be generalized to
\begin{equation}
\label{eq:Wilson-symmetry}
f(g(x))=U_{g}f(x)\,.
\end{equation}
The discrete symmetries $F^{S/R}$, which act freely on coset spaces $B_0=S/R$ are the center of $S$,
$\mathrm{Z}(S)$ and $\mathrm{W}=\mathrm{W}_{S}/\mathrm{W}_{R}$, where $\mathrm{W}_{S}$ and $\mathrm{W}_{R}$
are the Weyl groups of $S$ and $R$, respectively.  
The case of our interest here is
\begin{equation}
F^{S/R}= \mathbb{Z}_{3}\subseteq {\rm W}\, .
\end{equation}

\subsection{$SU(3)^3$ due to Wilson flux}

In order to derive the projected theory in the presence of the WL, one has to keep the
fields which are invariant under the combined action of the discrete group $\mathbb{Z}_3$ on the geometry and on the gauge indices.
The discrete symmetry acts non-trivially on the gauge fields and on the matter in the $27$ and the singlets. 
The action on the gauge indices is implemented via the matrix \cite{fuzzy}
${\rm diag}({\bf 1}_9,  \omega {\bf 1}_9,  \omega^2 {\bf 1}_9)$
with $\omega=e^{2i\pi/3}$. Thus, the gauge fields that survive the projection
are those that satisfy 
\begin{equation}
A_\mu = \gamma_3 A_\mu \gamma_3^{-1}\, ,\label{Aproj}
\end{equation}
while the surviving components of the matter fields in the 27's are those that satisfy
\begin{equation}
\vec \alpha = \omega \gamma_3 \vec\alpha\, , \hskip .5cm \vec\beta = \omega^2 \gamma_3 \vec\beta\, , \hskip .5cm \vec\gamma = \omega^3 \gamma_3 \vec\gamma\, .
\label{matterproj}
\end{equation}
Finally, the projection on the complex scalar singlets is
\begin{equation}
\alpha = \omega \alpha\, , \hskip .5cm \beta = \omega^2 \beta\, , \hskip .5cm \gamma = \omega^3 \gamma\, .
\end{equation}
It is easy to see then that after the $\mathbb{Z}_3$ projection the gauge group reduces to
\begin{equation}
A_\mu^A,\hskip 1cm A\in SU(3)_c\times SU(3)_L\times SU(3)_R
\end{equation}
(the first of the $SU(3)$ factors is the SM colour group)
and the scalar matter fields are in the bi-fundamental representations
\begin{equation}
\alpha_3 \sim {H_1} \sim ({\bar 3},1,{3})_{(3,1/2)}, \hskip .5cm 
\beta_2 \sim {H_2} \sim ({3},{\bar 3},1)_{(-3,1/2)}, \hskip .5cm
\gamma_1\sim {H_3} \sim (1,{3},{\bar 3})_{(0,-1)}.
\end{equation}
There are also fermions in similar representations. Note that with three
families this is a finite theory \cite{MaZoup,Mondragon1}.
Clearly, the Higgs is identified with the 9-component vector ${H_3}_a, a=1,\cdots ,9$.
Among the singlets, only ${\gamma}_{(0,-1)}$ survives.
In the following we will be using indices $a,b,c\cdots$ to count the complex components of a given
bi-fundamental representation and $i,j,k,\cdots = 1,2,3$ the different bifundamental representations.
 
Before we write the explicit scalar potential, we 
take appropriate actions such that there are 3 identical flavours from each of
the bifundamental fields. This can, in general, be achieved by introducing non-trivial windings in $R$.
We denote the resulting three copies of the bifundamental fields as (we will be using the index $l=1,2,3$ to specify the flavours)
\begin{eqnarray}
&& 3\cdot {H_1} \longrightarrow  {H_1}^{(l)} \sim 3\cdot  ({\bar 3},1,{3})_{(3,1/2)}  \nonumber\\
&& 3\cdot {H_2} \longrightarrow  {H_2}^{(l)}  \sim 3\cdot ({3},{\bar 3},1)_{(-3,1/2)}  \nonumber\\
&& 3\cdot {H_3} \longrightarrow  {H_3}^{(l)} \sim 3\cdot (1,{3},{\bar 3})_{(0,-1)}\, .
\end{eqnarray}
Similarly we denote the three copies of the scalar as
\begin{equation}
3\cdot {\gamma}_{{(0,-1)}}  \longrightarrow \theta^{(l)}_{{(0,-1)}}\, . 
\end{equation}
The scalar potential gets accordingly three copies of each contribution.

In the following when it does not cause confusion 
we denote a chiral superfield and its scalar component with the same letter. Also, it is clear that the potential after the 
projection will have the same form as before the projection
with the only difference that only $\theta^{(l)}$ is non-vanishing among the singlets and that the sums over 
components now run only over the even under the projection components.

We can now rewrite the scalar potential as \cite{Irges:2011de}
\begin{equation}
V_{\rm sc} = 3(3\Lambda^{2}+\Lambda'^{2})\biggl(\frac{1}{R_{1}^{4}}+\frac{1}{R_{2}^{4}}\biggr)
+\frac{3\cdot 4\Lambda'^{2}}{R_{3}^{4}} + \sum_{l=1,2,3} V^{(l)}
\end{equation}
where
\begin{equation}
V^{(l)} = V_{\rm susy} + V_{\rm soft}
\end{equation}
with $V_{\rm susy} = V_D + V_F$.
Since there are three identical contributions to the potential, at least until we give vevs to the Higgses
(which in general can be different for each $l$) we can drop the flavour superscript $(l)$ from most of the fields.
Then, the explicit form of the $D$ and $F$ terms are
\begin{eqnarray}
V_D &=& \frac{1}{2}\sum_A D^AD^A + \frac{1}{2}D_1D_1 + \frac{1}{2}D_2D_2 \nonumber\\
V_F &=&  \sum_{i=1,2,3} |F_{{H}_i}|^2 + |F_{\theta}|^2\, , \hskip .5cm F_{{H_i}} = \frac{\partial {\cal W}}{\partial {H_i}} ,  \hskip .5cm F_{{\theta}} = \frac{\partial {\cal W}}{\partial {\theta}}\, .
\end{eqnarray}
The $F$-terms derive from
\begin{equation}
{\cal W} = \sqrt{40} d_{abc} H_1^aH_2^bH_3^c
\end{equation}
and the $D$-terms are 
\begin{eqnarray}
D^A &=& \frac{1}{\sqrt{3}} \langle H_i | G^A | H_i \rangle \nonumber\\
D_1 &=& 3\sqrt{\frac{10}{3}}\left( \langle H_1 | H_1 \rangle -  \langle H_2 | H_2 \rangle \right)\nonumber\\
D_2 &=& \sqrt{\frac{10}{3}} \left( \langle H_1 | H_1 \rangle +  \langle H_2 | H_2 \rangle -2  \langle H_3 | H_3 \rangle - 2|\theta |^2 \right)\, ,
\end{eqnarray}
where 
\begin{eqnarray}
\langle H_i | G^A | H_i \rangle &=& \sum_{i=1,2,3}H_i^a (G^A)^b_a{H_i}_b \nonumber\\
\langle H_i | H_i \rangle &=&  \sum_{i=1,2,3} H_i^a\delta^b_a {H_i}_b\, .
\end{eqnarray}
Finally the soft breaking terms are
\begin{eqnarray}
V_{\rm soft} &=& \left(\frac{4R_1^2}{R_2^2R_3^2}-\frac{8}{R_1^2}\right) \langle H_1 | H_1 \rangle + \left(\frac{4R_2^2}{R_1^2R_3^2}-\frac{8}{R_2^2}\right) \langle H_2 | H_2 \rangle \nonumber\\
&+& \left(\frac{4R_3^2}{R_1^2R_2^2}-\frac{8}{R_3^2}\right) (\langle H_3 | H_3 \rangle +  |\theta |^2) \nonumber\\
&+& 80\sqrt{2} \left(\frac{R_1}{R_2R_3}+\frac{R_2}{R_1R_3} + \frac{R_3}{R_1R_2}\right)
(d_{abc} H_1^aH_2^bH_3^c + {\rm h.c.}).
\end{eqnarray}
The $(G^A)^b_a$ are structure constants, thus antisymmetric in $a$ and $b$.
The vector $|\phi \rangle$ and its hermitian conjugate $\langle \phi |$ represent the 9-dimensional bi-fundamental fields shown above.

The potential can be written in a more convenient form, as suggested in \cite{Kephart}. 
It amounts to writing the vectors in complex $3\times 3$ matrix notation. 
The various terms in the scalar potential can be then interpreted as invariant Lie algebra polynomials.
We identify
\begin{equation}
{H_1} \sim ({\bar 3},1,3) \longrightarrow {(q^c)}^\alpha_p\hskip .5cm
{H_2} \sim (3,{\bar 3},1) \longrightarrow Q^a_\alpha \hskip .5cm
{H_3} \sim (1,3,{\bar 3}) \longrightarrow L^p_a
\end{equation}
where
\begin{equation}
 q^c=\left(\begin{array}{ccc}
           d_R^1 & u_R^1 & D_R^1 \\
           d_R^2 & u_R^2 & D_R^2 \\
           d_R^3 & u_R^3 & D_R^3
           \end{array}
\right),\; 
Q=\left( 
\begin{array}{ccc}
d_L^1 & d_L^2 & d_L^3\\ u_L^1 & u_L^2 & u_L^3  \\ D_L^1 &  D_L^2 & D_L^3
\end{array}
\right), \; 
L=\left(
\begin{array}{ccc}
 H_d^0 & H_u^+ & v_L \\
 H_d^- & H_u^0 & e_L \\
v_R & e_R & S
\end{array}
\right)\;.
\end{equation}
Evidently $d_{L,R}, u_{L,R}, D_{L,R}$ tranforms as $3,\bar 3$ under colour. Then
we
introduce
\begin{equation}
{\hat{q^c}}_\alpha^p = \frac{1}{3} \frac{\partial I_3}{\partial
{q^c}^\alpha_p}\hskip .5cm
{\hat L}_p^a = \frac{1}{3} \frac{\partial I_3}{\partial L^p_a}\hskip .5cm
{\hat Q}_a^\alpha = \frac{1}{3} \frac{\partial I_3}{\partial Q^a_\alpha}\, ,
\end{equation}
where
\begin{equation}
I_3 = {\rm det [Q]} + {\rm det [q^c]} + {\rm det [L]} -{\rm tr}(q^c\cdot L\cdot
Q )\, .
\end{equation}
In terms of these matrices, we have
$\langle H_1|H_1\rangle = {\rm tr} ({q^c}^\dagger {q^c})$, 
$\langle H_2|H_2\rangle = {\rm tr} (Q^\dagger Q)$,
$\langle H_3|H_3\rangle = {\rm tr} (L^\dagger L)$
and 
\begin{equation}
d_{abc} H_1^aH_2^bH_3^c = {\rm det {q^c}^\dagger} + {\rm det M^\dagger} + {\rm
det L^\dagger} -{\rm tr}(N^\dagger M^\dagger L^\dagger)\, .
\end{equation}
The $F$-terms which explicitly read
\begin{eqnarray}
V_F &=& 40 d_{abc}d^{cde} ({H_1}^a {H_2}^b{H_1}_d{H_2}_e + {H_2}^a {H_3}^b{H_2}_d{H_3}_e + {H_1}^a {H_3}^b{H_1}_d{H_3}_e).
\end{eqnarray}
can be now written as
\begin{equation}
V_F = 40 {\rm tr} ({\hat {q^c}}^\dagger {\hat {q^c}}+{\hat Q}^\dagger {\hat
Q}+{\hat L}^\dagger {\hat L}) .
\end{equation}

\section{Gauge symmetry breaking}
 
Consider the following vevs:
\begin{equation}
L_0^{(1)}  = \begin{pmatrix} 
0 & 0 & 0 \cr 
0 & 0 & 0 \cr
0 & 0 & V
\end{pmatrix}, \hskip 1cm
L_0^{(2)}  = \begin{pmatrix} 
0 & 0 & 0 \cr 
0 & 0 & 0 \cr
V & 0 & 0
\end{pmatrix}\label{vev2}
\end{equation}
for $H_3^{(1)}$ and $H_3^{(2)}$ respectively. These vevs leave the $SU(3)_c$
part of the gauge group unbroken
but trigger the spontaneous breaking of the rest.
More precisely, $L_0^{(1)}$ breaks the gauge group according to
\begin{equation}
SU(3)_c\times SU(3)_L\times SU(3)_R \times U(1)_A \times U(1)_B \longrightarrow
SU(3)_c\times SU(2)_L\times SU(2)_R\times U(1)
\end{equation}
while $L_0^{(2)}$ according to
\begin{equation}
SU(3)_c\times SU(3)_L\times SU(3)_R \times U(1)_A \times U(1)_B\longrightarrow 
SU(3)_c\times SU(2)_L\times SU(2)_R'\times U(1)' \, .
\end{equation}
The combination of the two gives \cite{Will}
\begin{equation}
SU(3)_c\times SU(3)_L\times SU(3)_R  \times U(1)_A \times U(1)_B \longrightarrow
SU(3)_c\times SU(2)_L\times U(1)_Y\, .
\end{equation}
Electroweak (EW) breaking then proceeds by a second vev $v$, for example by
\cite{MaZoup}
$L_{0}^{(1)} = {\rm diag}(v,v,V)$.
We first look at $V^{(1)}$ in the presence of the vevs.
Using the fact that the coefficients ${(G^A)}^b_a$ are antisymmetric in $a$ and
$b$,
it is easy to see that for these vevs, the 
quadratic form $\langle  \phi | G^A | \phi \rangle$ vanishes identically in the
vacuum, 
and so do the corresponding $SU(3)$ $D$-terms $D^A$.
The other terms give in the vacuum
\begin{eqnarray}
V_{D_1} &=& 15 (V^2+2v^2)^2 \nonumber\\
V_{D_2} &=& \frac{5}{9} (V^2+2v^2-\theta_0^2)^2 \nonumber\\
V_F &=& \frac{40}{9} v^2(2 V^2 + v^2) \nonumber\\
V_{\rm soft} &=& \left(\frac{4R_2^2}{R_1^2R_3^2}-\frac{8}{R_2^2}\right) (V^2 +
2v^2) +
\left(\frac{4R_3^2}{R_1^2R_2^2}-\frac{8}{R_3^2}\right) (\theta^{(1)}_0)^2
\nonumber\\
&+&160\sqrt{3} 
\left(\frac{R_1}{R_2R_3}+\frac{R_2}{R_1R_3} + \frac{R_3}{R_1R_2}\right) V v^2 \,
.
\end{eqnarray} 
As expected, already in the vacuum where EW symmetry is unbroken, supersymmetry
is broken by both
$D$ and $F$-terms, in addition to its breaking by the soft terms.  
The potential is positive definite so we are looking for a 
vacuum solution with $V^{(1)}_0=0$.  
For simplicity we choose $R_1=R_2=R_3=R$
(strictly speaking in this case the manifold becomes nearly-K\"ahler). Then, if
the vevs satisfy the relation
\begin{eqnarray}
(\theta^{(1)}_0)^2 &=& \frac{1}{10R^2} \Bigl[5R^2V^2 + 10R^2v^2 + 9 \nonumber\\
&+& \bigl( -675V^4R^4 -3100 V^2v^2 R^4 +270 V^2 R^2 -2900 v^4R^4 \nonumber\\
&+& 540 v^2R^2 
+27-21600\sqrt{3} Vv^2 R^3\bigr)^{1/2}
\Bigr] \, ,
\end{eqnarray}
the potential is zero at the minimum. We stress that in contrast to exactly
supersymmetric theories, the 
zero of the potential at the minimum does not imply unbroken supersymmetry. This
is because the
potential is a perfect square (which is a consequence of its higher dimensional
origin from $F_{MN}F^{MN}$) with
the soft breaking terms included.  

It is interesting to notice that generically the solution makes sense when $V <
1/R$
and then if we set, with no loss of generality $V=1$, we find that the quantity
under the square root is positive
if $v \sim {\cal O}( 0.1)$ for $R\sim {\cal O} (1/2)$. 
It is interesting that the desired hierarchy of scales is naturally generated by
the 
structure of the scalar potential. Therefore a consistent picture emerges
assuming that the compactification
and the supersymmetry breaking scales are both in the few TeV regime
\cite{AntoniadisI}.

The analysis of $V^{(2)}$ in the presence of the second of the vevs in eq.
(\ref{vev2}) is similar. The potential 
is zero at the minimum if the vev of $\theta^{(2)}$ satisfies
\begin{equation}
(\theta^{(2)}_0)^2 = \frac{1}{10 R^2}\left(5V^2R^2 +9+
3\sqrt{-75V^4R^4+30V^2R^2+3}\right).
\end{equation}
The vevs $\theta^{(1)}_0$ and $\theta^{(2)}_0$ need not be equal and
$\theta^{(3)}_0$ can, but does not need 
to be zero.

\subsection{$U(1)$ structure and Yukawa couplings}

The breaking pattern of the bifundamental representations  that $V$ induces  is 
\begin{eqnarray}
&&({\overline 3},1,3)_{(3,1/2)} \longrightarrow ({\overline
3},1,1+1+1)_{(3,1/2)}\label{rep1}\\
&&(3,{\overline 3},1)_{(-3,1/2)}\longrightarrow
(3,2+1,1)_{(-3,1/2)}\label{rep2}\\
&&(1,3,{\overline 3})_{(0,-1)}\longrightarrow (1,2+1,1+1+1)_{(0,-1)}\label{rep3}
\end{eqnarray}
from which we can read off the representations under the SM gauge group and the
extra
$U(1)$'s. From (\ref{rep1}) we obtain ${\bf \overline u}$, ${\bf \overline d}$
and ${\bf \overline D}$,
that is the two right handed quarks and an extra quark type state. From
(\ref{rep2}) we obtain 
the quark doublet ${\bf Q}$ and the vector-like partner ${\bf D}$ of the extra
quark.
Notice however that the extra quarks are not completely vector-like, since they
have the same 
$U(1)_B$ charge. From (\ref{rep3}) we obtain the lepton doublet $L$, the right
handed lepton
singlet ${\overline e}$, two right handed neutrinos and two electroweak
doublets.
We will denote the latter doublets as $H_u$ and $H_d$ like in the minimal
supersymmetric SM (MSSM). 
Notice that the scalar components of these doublets are the components of the
$H_3$ Higgs field
that takes the vev $v$.
We will denote the former singlets as ${\overline N}_{1,2}$ 
while the singlet chiral superfields whose lowest component are the
$\theta^{(l)}$
we will call ${\Theta}^{(l)}$.  
In the table we summarize the states contained in one family, with their $U(1)$
charges.
We have separated the MSSM spectrum from new states by a double line.
 
\vskip 0.2cm
\hskip 2cm
\begin{center}
\begin{tabular}{|c|c|c|c|}
\hline 
$ SU(3)_c\times SU(2)_L $ & $U(1)_Y$ & $ U(1)_A $ & $ U(1)_B $ \\
\hline \hline   
${\bf Q}\sim (3,2)$ & $1/6$ &$-3$ & $1/2$\\ \hline
${\bf \overline u}\sim ({\bar 3},1)$ & $-2/3$ & $3$ & $1/2$\\ \hline            
                    
${\bf \overline d}\sim ({\bar 3},1) $ & $1/3$ & $3$ & $1/2$ \\ \hline
$ L \sim (1,2)$ & $-1/2$ & $0$ & $-1$ \\ \hline
${\overline e}\sim (1,1)$ & $1$ & $0$ & $-1$ \\ \hline
$H_u\sim (1,2)$ & $1/2$ & $0$ & $-1$ \\ \hline
$H_d\sim (1,2)$ & $-1/2$ & $0$ & $-1$ \\ \hline\hline
${\bf D}\sim (3,1)$ & $-1/3$ & $-3$ & $1/2$  \\ \hline
${\bf \overline D}\sim ({\bar 3},1)$ & $1/3$ & $3$ & $1/2$ \\ \hline
${\overline N_1}\sim (1,1)$ & $0$ & $0$ & $-1$ \\ \hline
${\overline N_2}\sim (1,1)$ & $0$ & $0$ & $-1$ \\ \hline
${\Theta}^{(1)}\sim (1,1)$ & $0$ & $0$ & $-1$ \\ \hline
\end{tabular}
\end{center}
\vskip 0.3cm
We immediately recognize that
\begin{equation}
U(1)_A = -9 B\, ,
\end{equation}
where $B$ is baryon number and $U(1)_B$ as a Peccei-Quinn type of symmetry.
Lepton number on the other hand does not appear to be a conserved symmetry (e.g.
$L{\bf Q}{\bf \overline d}$ is allowed).
The presence of a conserved global baryon number is clearly a welcome feature
from the point of view of
the stability of the proton.
The two extra $U(1)$'s at this stage, are both anomalous and at least one of
them will remain anomalous after charge
redefinitions. 
They both break by the vev $V$, however their respective global subgroups remain
at low energies and constrain
the allowed (non-renormalizable) operators in the superpotential.
Gauge invariance in the presence of the anomalous symmetries can be maintained
by the addition of a specific combination of terms
to the low energy effective Lagrangian,
including a St\"uckelberg coupling and an axion-like interaction. 
These interactions introduce a new, phenomenologically interesting sector in the
effective action \cite{cen}.
Let $A_\mu$ be the anomalous $U(1)$ gauge field and $F_A$ its field strength.
Then the terms that render the
action gauge invariant are
\begin{equation}
{\cal L}_{\rm St-WZ} = \frac{1}{2}(\partial_\mu a + M A_\mu)^2 + c \frac{a}{M}
F_A\wedge F_A + 
{\cal L}_{\rm an}\, .
\end{equation}
The axion $a$ shifts under the anomalous symmetry so that the kinetic term is
invariant
and coefficient $c$ is such that the Wess-Zumino term cancels the 1-loop anomaly
${\cal L}_{\rm an}$.
The scale $M$ is related to the vev $V$. 
These couplings are added by hand because they are not part of the 
the interactions of the original ten-dimensional gauge multiplet, neither can be
generated by its
dimensional reduction. In fact, the axion field is the four-dimensional remnant
of the two-form $B_{MN}$.
This is the (minimum) price to pay for neglecting the gravitational and two-form
sectors 
(and actually also the second $E_8$ factor) along with the ten-dimensional
anomaly cancellation mechanism,
for which their presence is essential \cite{GS}.

A few comments regarding the Yukawa sector are in order. Every operator
originating from
the superpotential $d_{abc}H_1^aH_2^bH_3^c$ will appear at tree level.
At the quantum level operators that break the CSDR constraints and the
supersymmetric structure will eventually 
develop, as long as they are gauge invariant. As an example of the former case,
the extra vector-like pair of quarks will develop a mass term in the $V$-vacuum
\begin{equation}
\Theta^{(1)} {\bf \overline D}{\bf D}
\end{equation}
which is a singlet. As an example of the latter, notice that in the quark sector
the standard Yukawa terms in the superpotential appear at tree level. 
In the lepton sector however the term $L{\overline e}H_d$ is not invariant under
$U(1)_B$. 
An effective Yukawa coupling can come though from the higher-dimensional
operator
\begin{equation}
L{\overline e} H_d \left( \frac{{\theta^{(1) *}}}{M}\right)^3
\end{equation}
in the $V$-vacuum, with $M$ a high scale such as the string scale and
$\theta^{(1)*}$ the complex conjugate of $\theta^{(1)}$. 
Similar arguments apply to the entire lepton sector:
effective Yukawa couplings appear via higher dimensional operators
\begin{equation}
L H_u {\overline N} \left( \frac{{\theta^{(1)*}}}{M}\right)^3\hskip 1cm
M  {\overline N} {\overline N}  \left( \frac{{\theta^{(1)*}}}{M}\right)^2\, .
\end{equation}
Similar terms are generated for the second and third families.
Evidently, after electroweak symmetry breaking, fermion mass hierarchies and
mixings can be generated \cite{FN}
not because the $U(1)$'s have  flavour dependent charges,
but from the different values that the vevs $\theta^{(l)}$ can have. 
A term that mixes flavours is, for example,
\begin{equation}
L^{(1)}{\overline e}^{(2)} H_d^{(2)} \left( \frac{{\theta^{(1)
*}}}{M}\right)\left( \frac{{\theta^{(2) *}}}{M}\right)^2\, ,
\end{equation}
where we have made the flavour superscripts explicit on all fields.

Clearly the general flavour problem is a crucial question to address in this
model.
The symmetries of the coset space are very constraining on the 
four-dimensional effective action and so will be the quark and lepton mass
hierarchies. It will be therefore interesting to see if the observed pattern
in the mass hierarchies and mixings is possible to be accommodated.
In a future work, we plan to investigate this important issue.

\vskip 1cm
{\bf Acknowledgements}

This work was partially supported by the NTUA's basic 
research support programmes PEVE2009 and PEVE2010 and the European Union ITN programme "UNILHC" PITN-GA-2009-237920.
G. O. and G. Z. would like to thank CERN for hospitality.

\newpage
\appendix
\section*{Appendices}
\section{Reducing the 10-dimensional 32-spinor to 8-spinor by Majorana-Weyl
conditions
}\label{appendix}

The case we are going to examine is ${\cal N}=1$ SYM theory in D=10.  In
particular we would like to demonstrate how the 
Dirac spinor with $2^{[D/2]}=32$ components is reduced to a 
Weyl-Majorana spinor with 8 components in order to
have the same degrees of freedom as the gauge fields.
We choose the following representation of the 
 $\Gamma$-matrices 
\begin{equation}
 \Gamma^\mu=\gamma^\mu \otimes I_8\;,\; \mu=0,1,2,3\;\; .
\end{equation}
 The Dirac spinor can be written as
\begin{equation}
 \psi=\left(\psi_1\; \dots\; \psi_4 \;\chi_1 \;\dots \chi_4 \;\right)^T\;,
\end{equation}
where all $\psi_i,\;\chi_i,\;\;i=1,\dots,4$ tranform as $SO(1,3)$ Dirac
spinors. Let us present the rest $\Gamma$-matrices
\begin{eqnarray}
 \Gamma^4=\gamma^5\otimes \sigma^1 \otimes \sigma^2 \otimes \sigma^2 &,&
\Gamma^5=\gamma^5\otimes \sigma^2\otimes \sigma^2 \otimes \sigma^2,\nonumber\\ 
\Gamma^6=\gamma^5\otimes I_2 \otimes \sigma^3 \otimes
\sigma^2 & , & \;\Gamma^7=\gamma^5\otimes I_4 \otimes
\sigma^1,\nonumber\\ 
\Gamma^8=\gamma^5\otimes \sigma^3 \otimes \sigma^2 \otimes
\sigma^2 & , & \;\Gamma^9=\gamma^5\otimes I_2
\otimes \sigma^1 \otimes \sigma^2
\end{eqnarray}
and hence
\begin{equation}
 \Gamma^{11}=\Gamma^0\dots\Gamma^9=-\gamma^5\otimes I_2\otimes I_2 \otimes
\sigma^3=\gamma^5\otimes
\left(\begin{array}{ccc}
       -I_4 & 0\\
       0   & I_4
      \end{array}
\right)
\end{equation}
The spinor $\psi$ is reducible, $\Gamma^{11}\psi_\pm=\pm\psi_\pm$, where
$\psi_\pm=\frac{1}{2}\left(1\pm\Gamma^{11}\right)\psi$. 
Then the Weyl condition,
\begin{equation}
\Gamma^{11} \psi = \psi
\end{equation}
selects the $\psi_+$ ,
where
\begin{equation}
  \psi_+=\left(L\psi_1\; \dots\; L\psi_4 \;R\chi_1 \;\dots R\chi_4
\;\right)^T
\end{equation}
where $L=\frac{1}{2}\left(1-\gamma^5\right)$ (left handed) and
$R=\frac{1}{2}\left(1+\gamma^5\right)$ (right handed).
 The $\psi_i$ form  the 
 4  and the $\chi_i$  the  $\bar{4}$
representations of $SO(6)$. Imposing further the Majorana condition on the
10-dimensional spinor,
\begin{equation}
 \psi=C_{10}\Gamma^0\psi^*
\end{equation}
where $C_{10}=C_4\otimes\sigma_2\otimes\sigma_2\otimes I_2$
we are led to the relations $\chi_{1,3}=C\gamma_0\psi^*_{2,4}$ and
$\chi_{2,4}=-C\gamma_0\psi^*_{1,3}$. Therefore by imposing both Weyl and
Majorana conditions in 10 dimensions we obtain a Weyl spinor in 4 dimensions
transforming as $4$ of $SO(6)$ i.e.
\begin{equation}
\psi= \left(L\psi_1\;\;L\psi_2\;\;L\psi_3 \;\; L\psi_4 \;\;R\tilde{\psi}_2
\;\;R\tilde{\psi}_1\;\;R\tilde{\psi}_4\;\; R\tilde{\psi}_3
\;\right)^T\;,\; \tilde{\psi}_i=(-1)^iC\gamma_0\psi^*_i\;.
\end{equation}

In addition we
need the gamma matrices in the coset space $SU(3)/U(1)\times U(1)$. The metric
is given 
$g_{ab}=diag(R_1^2,R_1^2,R_2^2,R_2^2,R_3^2,R_3^2)$, $g^{ab}=diag(\frac{1}{
R_1^2 } , \frac{1}{R_1^2} , \frac{1}{R_2^2}
,\frac{1}{R_2^2},\frac{1}{R_3^2},\frac{1}{R_3^2})$ and hence the
$\Gamma$-matrices are given by
\begin{eqnarray}
 \Gamma^4=\frac{1}{R_1}\gamma^5\otimes \sigma^1 \otimes \sigma^2 \otimes
\sigma^2 & , & \Gamma^5=\frac{1}{R_1}\gamma^5\otimes \sigma^2
\otimes \sigma^2 \otimes \sigma^2,\nonumber\\ 
\Gamma^6=\frac{1}{R_2}\gamma^5\otimes I_2 \otimes \sigma^3 \otimes
\sigma^2 & , &
\Gamma^7=\frac{1}{R_2}\gamma^5\otimes I_4 \otimes
\sigma^1,\nonumber\\ 
\Gamma^8=\frac{1}{R_3}\gamma^5\otimes \sigma^3 \otimes \sigma^2 \otimes
\sigma^2 & , &
\Gamma^9=\frac{1}{R_3}\gamma^5\otimes I_2
\otimes \sigma^1 \otimes \sigma^2
\end{eqnarray}

\section{Commutation Relations}\label{tables}

\begin{table}[h]
\caption{}
\centering
\begin{tabular}{|l l l|} \hline
 \multicolumn{3}{|c|}{
Six-dimensional non-symmetric cosets with $rankS=rankR$}\\ \hline
$S/R$ & $SO(6)$ vector & $SO(6)$ spinor \\ \hline $G_{2}/SU(3)$ &
$3+\overline{3}$ & $1+3$ \\ $Sp(4)/(SU(2) \times U(1))_{non-max}$
& $1_{2}+1_{-2}+2_{1}+2_{-1}$ & $1_{0}+1_{2}+2_{-1}$ \\
$SU(3)/U(1) \times U(1)$ & $(a,c)+(b,d)+(a+b,c+d)$ &
$(0,0)+(a,c)+(b,d)$ \\ & $+(-a,-c)+(-b,-d)$ & $+(-a-b,-c-d)$ \\ &
$+(-a-b,-c-d)$ &  \\ \hline
\end{tabular} \\ \vspace{.2in}
\label{table4}
\end{table}

\begin{table}[h]
\caption{}
\centering
\begin{tabular}{|l|l|l|}\hline
\multicolumn{3}{|c|}{Non-trivial
commutation relations of $SU(3)$ according to} \\
\multicolumn{3}{|c|}{the decomposition given in \eqref{pan:80}}\\ \hline
$\left[Q_{1},Q_{0}\right]=\sqrt{3}Q_{1}$ &
$\left[Q_{1},Q'_{0}\right]=Q_{1}$ &
$\left[Q_{2},Q_{0}\right]=-\sqrt{3}Q_{2}$\\
$\left[Q_{2},Q'_{0}\right]=Q_{2}$ & $\left[Q_{3},Q_{0}\right]=0$ &
$\left[Q_{3},Q'_{0}\right]=-2Q_{3}$ \\
$\left[Q_{1},Q^{1}\right]=-\sqrt{3}Q_{0}-Q'_{0}$ &
$\left[Q_{2},Q^{2}\right]=\sqrt{3}Q_{0}-Q'_{0}$ &
$\left[Q_{3},Q^{3}\right]=2Q'_{0}$ \\
$\left[Q_{1},Q_{2}\right]=\sqrt{2}Q^{3}$ &
$\left[Q_{2},Q_{3}\right]=\sqrt{2}Q^{1}$ &
$\left[Q_{3},Q_{1}\right]=\sqrt{2}Q^{2}$\\ \hline
\end{tabular}
\label{table1}
\end{table}
The normalization in the above table is
$$Tr(Q_{0}Q_{0})=Tr(Q'_{0}Q'_{0})=Tr(Q_{1}Q^{1})=Tr(Q_{2}Q^{2})=Tr(Q_{3}Q^{3}
)=2$$

\begin{table}[h]
\caption{}
\centering
\begin{tabular}{|l|l|l|}\hline
\multicolumn{3}{|c|}{
Non-trivial commutation relations of $E_{8}$ according to} \\
\multicolumn{3}{|c|}{the decomposition given by eq.\eqref{pan:84}}\\ \hline
$\left[Q_{1},Q_{0}\right]=\sqrt{30}Q_{1}$ &
$\left[Q_{1},Q'_{0}\right]=\sqrt{10}Q_{1}$ &
$\left[Q_{2},Q_{0}\right]=-\sqrt{30}Q_{2}$ \\
$\left[Q_{2},Q'_{0}\right]=\sqrt{10}Q_{2}$ &
$\left[Q_{3},Q_{0}\right]=0$ &
$\left[Q_{3},Q'_{0}\right]=-2\sqrt{10}Q_{3}$ \\
$\left[Q_{1},Q^{1}\right]=-\sqrt{30}Q_{0}-\sqrt{10}Q'_{0}$ &
$\left[Q_{2},Q^{2}\right]=\sqrt{30}Q_{0}-\sqrt{10}Q'_{0}$ &
$\left[Q_{3},Q^{3}\right]=2\sqrt{10}Q'_{0}$ \\
$\left[Q_{1},Q_{2}\right]=\sqrt{20}Q^{3}$ &
$\left[Q_{2},Q_{3}\right]=\sqrt{20}Q^{1}$ &
$\left[Q_{3},Q_{1}\right]=\sqrt{20}Q^{2}$ \\
$\left[Q_{1i},Q_{0}\right]=\sqrt{30}Q_{1i}$ &
$\left[Q_{1i},Q'_{0}\right]=\sqrt{10}Q_{1i}$ &
$\left[Q_{2i},Q_{0}\right]=-\sqrt{30}Q_{2i}$ \\
$\left[Q_{2i},Q'_{0}\right]=\sqrt{10}Q_{2i}$ &
$\left[Q_{3i},Q_{0}\right]=0$ &
$\left[Q_{3i},Q'_{0}\right]=-2\sqrt{10}Q_{3i}$ \\
$\left[Q_{1i},Q_{2j}\right]=\sqrt{20}d_{ijk}Q^{3k}$ &
$\left[Q_{2i},Q_{3j}\right]=\sqrt{20}d_{ijk}Q^{1k}$ &
$\left[Q_{3i},Q_{1j}\right]=\sqrt{20}d_{ijk}Q^{2k}$ \\
$\left[Q^{\alpha},Q^{\beta}\right]=2ig^{\alpha\beta\gamma}Q^{\gamma}$
& $\left[Q^{\alpha},Q_{0}\right]=0$ &
$\left[Q^{\alpha},Q'_{0}\right]=0$ \\
$\left[Q^{\alpha},Q_{1i}\right]=-(G^{\alpha})^{j}_{i}Q_{1j}$ &
$\left[Q^{\alpha},Q_{2i}\right]=-(G^{\alpha})^{j}_{i}Q_{2j}$ &
$\left[Q^{\alpha},Q_{3i}\right]=-(G^{\alpha})^{j}_{i}Q_{3j}$\\
\hline \multicolumn{3}{|c|}{
$\left[Q_{1i},Q^{1j}\right]=-\frac{1}{6}(G^{\alpha})^{j}_{i}Q^{\alpha}
-\sqrt{30}\delta^{j}_{i}Q_{0}-\sqrt{10}\delta^{j}_{i}Q'_{0}$ }\\
\multicolumn{3}{|c|}{
$\left[Q_{2i},Q^{2j}\right]=-\frac{1}{6}(G^{\alpha})^{j}_{i}Q^{\alpha}
+\sqrt{30}\delta^{j}_{i}Q_{0}-\sqrt{10}\delta^{j}_{i}Q'_{0}$ }\\
\multicolumn{3}{|c|}{
$\left[Q_{3i},Q^{3j}\right]=-\frac{1}{6}(G^{\alpha})^{j}_{i}Q^{\alpha}
+2\sqrt{10}\delta^{j}_{i}Q'_{0}$}\\ \hline
\end{tabular}
\label{table2}
\end{table}

The normalization in the table \ref{table2}  is
\begin{eqnarray*}
Tr(Q_{0}Q_{0})&=&Tr(Q'_{0}Q'_{0})=Tr(Q_{1}Q^{1})=Tr(Q_{2}Q^{2}
)=Tr(Q_ { 3 } Q^ { 3 }
)=2\;, \\
Tr(Q_{1i}Q^{1j})&=&Tr(Q_{2i}Q^{2j})=Tr(Q_{3i}Q^{3j})=2\delta^{j}_{i},
\;\;Tr(Q^{\alpha }Q^{\beta})=12\delta^{\alpha\beta}.
\end{eqnarray*}

\section{Useful relations to calculate the gaugino mass.\label{appC}} 
We use the real metric of the coset,
$g_{ab}=diag(a,a,b,b,c,c)$ with
$a=R_{1}^{2},b=R_{2}^{2},c=R_{3}^{2}$. Using the structure
constants of $SU(3)$, $ f_{12}^{\ \ 3}=2 $, $ f_{45}^{\ \
8}=f_{67}^{\ \ 8}=\sqrt{3} $, $ f_{24}^{\ \ 6} =f_{14}^{\ \
7}=f_{25}^{\ \ 7}=-f_{36}^{\ \ 7}=-f_{15}^{\ \ 6}=-f_{34}^{\ \
5}=1$, (where the indices 3 and 8 correspond to the $U(1) \times
U(1)$ and the rest are the coset indices) we calculate the
components of the $D_{abc}$:\\ $D_{523}
=D_{613}=D_{624}=D_{541}=-D_{514}=-D_{532}=-D_{631}=-D_{624}=\frac{1}{2}
(c-a-b).$\\
$D_{235}=D_{136}=D_{624}=D_{154}=-D_{145}=-D_{253}=-D_{163}=-D_{264}=\frac{1}{2}
(a-b-c).$\\
$D_{352}=D_{361}=D_{462}=D_{415}=-D_{451}=-D_{325}=-D_{316}=-D_{
426}=\frac{1}{2}(b-c-a).$\\ From the $D$'s we calculate the
contorsion tensor $$ \Sigma_{abc}=2\tau(D_{abc}+D_{bca}-D_{cba}),
$$ and then the tensor $$G_{abc}=D_{abc}+\frac{1}{2}\Sigma_{abc}$$
which is\\
{\small
$G_{523}=G_{613}=G_{642}=G_{541}=-G_{514}=-G_{532}=-G_{631}=-G_{642}=
\frac{1}{2}[(1-\tau)c-(1+\tau)a-(1+\tau)b].$\\
$G_{235}=G_{136}=G_{246}=G_{154}=-G_{145}=-G_{253}=-G_{163}=-G_{264}=
\frac{1}{2}[-(1-\tau)a+(1+\tau)b+(1+\tau)c].$\\
$G_{352}=G_{361}=G_{462}=G_{415}=-G_{451}=-G_{325}=-G_{316}=-G_{426}=
\frac{1}{2}[-(1+\tau)a+(1-\tau)b-(1+\tau)c].$
}

\end{document}